\documentclass[reprint,showpacs,preprintnumbers,amsmath,amssymb,aps,prl]{revtex4-1}
\usepackage{graphicx}
\usepackage{dcolumn}
\usepackage{bm}
\usepackage{multirow}
\usepackage{bbding}
\usepackage{color}
\usepackage{ulem}

\allowdisplaybreaks[1]

\begin{document}
	
	\title{Spin elastodynamic motive force}
	\author{Takumi Funato${}^{1,2}$ and Mamoru Matsuo${}^{2,3,4,5}$}
	\affiliation{${}^1$Center for Spintronics Research Network, Keio University, Yokohama 223-8522, Japan}
	\affiliation{%
		${}^2$Kavli Institute for Theoretical Sciences, University of Chinese Academy of Sciences, Beijing, 100190, China.
	}%
	\affiliation{%
		${}^3$CAS Center for Excellence in Topological Quantum Computation, University of Chinese Academy of Sciences, Beijing 100190, China
	}%
	\affiliation{${}^4$RIKEN Center for Emergent Matter Science (CEMS), Wako, Saitama 351-0198, Japan}
	\affiliation{
		${}^5$Advanced Science Research Center, Japan Atomic Energy Agency, Tokai, 319-1195, Japan
	}
	
	\date{\today}
	
	\begin{abstract}
		The spin-motive force (SMF) in a simple ferromagnetic monolayer caused by a surface acoustic wave is studied theoretically via spin-vorticity coupling (SVC). 
		The SMF has two mechanisms. The first is the SVC-driven SMF, which produces the first harmonic electromotive force, and the second is the interplay between the SVC and the magentoelastic coupling, which produces the d.c. and second harmonic electromotive forces.  
		We show that these electric voltages induced by a Rayleigh-type surface acoustic wave can be detected in polycrystalline nickel.
		No sophisticated device structures, non-collinear magnetic structures, or strong spin-orbit materials are used in our approach.
		Consequently, it is intended to broaden the spectrum of SMF applications considerably.
	\end{abstract}
	
	\pacs{Valid PACS appear here}
	\maketitle
	
	
	{\it Introduction}.--- Spintronics is concerned with the interconversion of charge transport and spin dynamics.
	The use of charge current to control magnetization structures, such as magnetic domain wall, has been extensively investigated for spin-torque oscillators\cite{kiselevMicrowaveOscillationsNanomagnet2003,rippardDirectCurrentInducedDynamics2004,katineDeviceImplicationsSpintransfer2008} and magnetoresistive random access memory applications\cite{hosomiNovelNonvolatileMemory2005,ikedaMagneticTunnelJunctions2007,matsunagaFabricationNonvolatileFull2008a}.
	The spin-transfer torque\cite{slonczewskiCurrentdrivenExcitationMagnetic1996,bergerEmissionSpinWaves1996,katineCurrentDrivenMagnetizationReversal2000} associated with adiabatically following conduction electron spins to the magnetization through the s-d coupling plays a significant role.
	The spin motive force (SMF), a current-driven phenomenon related to the spin-dependent force, was theoretically hypothesized as an inverse spin-transfer torque effect\cite{bergerPossibleExistenceJosephson1986,volovikLinearMomentumFerromagnets1987,sternBerryPhaseMotive1992,barnesGeneralizationFaradayLaw2007}.
	The first experiment involves observing the SMF in a nanowire due to magnetic domain wall motion\cite{yangUniversalElectromotiveForce2009,yangTopologicalElectromotiveForce2010,hayashiTimeDomainObservationSpinmotive2012}.
	
	In nonuniform magnetic textures formed by a comb-shaped device\cite{yamaneContinuousGenerationSpinmotive2011}, the magnetic vortex on a gyrating disk\cite{tanabeSpinmotiveForceDue2012a}, a wedged-shaped device\cite{nagataSpinMotiveForce2015}, exchange-coupled ferromagnetic bilayers\cite{zhouSpinmotiveForceOutofplane2019}, and helical magnetism\cite{nagaosaEmergentInductorSpiral2019,yokouchiEmergentElectromagneticInduction2020}, the SMF induced by excited magnetization dynamics was seen.
	The SMF induced by the oscillating magnetic field in systems with Rashba spin-orbit interaction (SOI)\cite{kimPredictionGiantSpin2012} and temporarily varying gate voltage in systems with strong SOI\cite{hoSustainableSpinCurrent2012,hoSpinForceGeneration2014,hoGatecontrolSpinmotiveForce2015,yamaneSpinmotiveForceStatic2013} have not been observed experimentally.
	There is little variety in experimental reports on the SMFs because of experimental restrictions such as sophisticated device structures, non-collinear magnetic structures, and strong SOI systems.
	
		\begin{figure}[tbp]
		\centering
		\includegraphics[width=75mm]{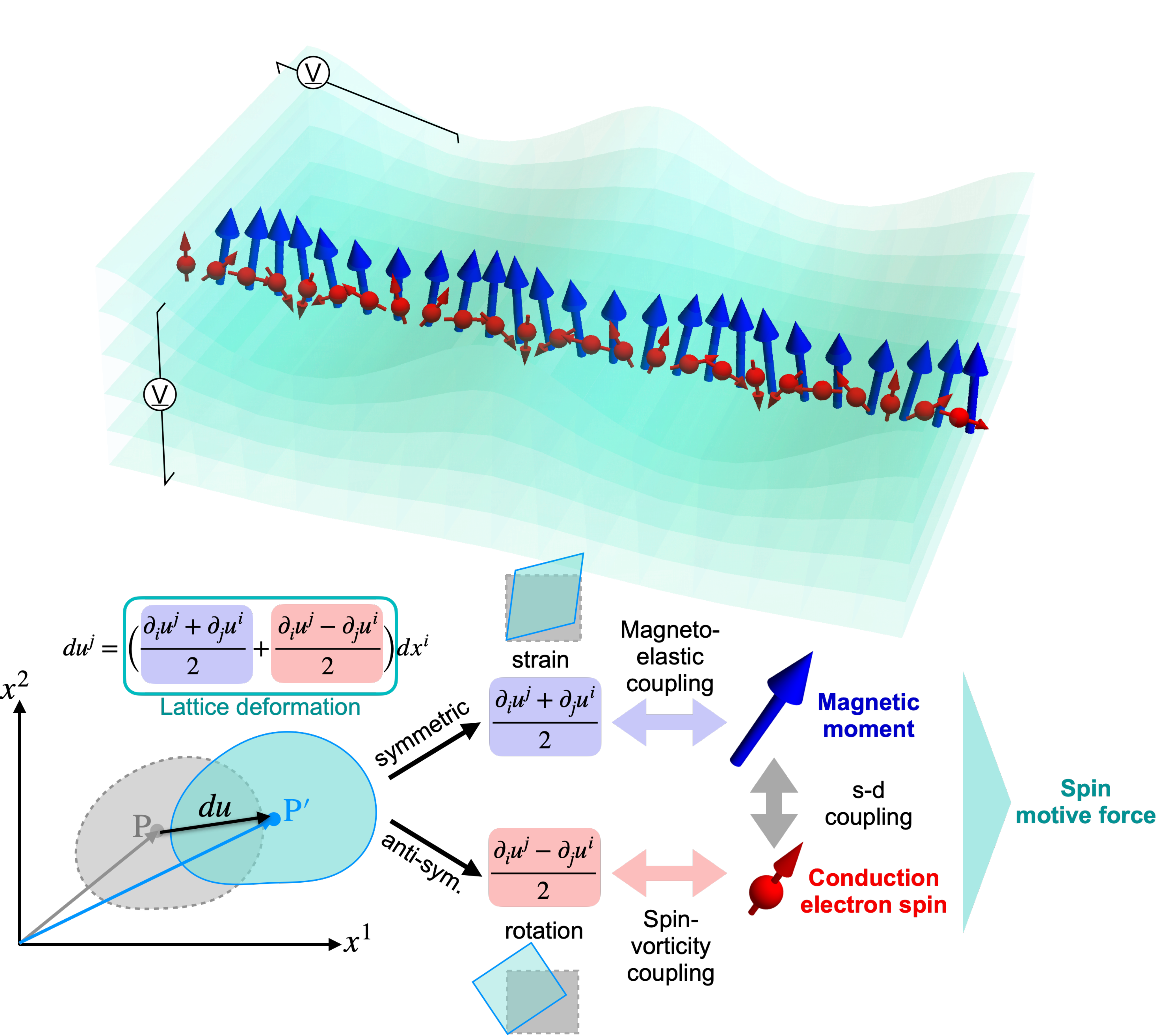}
		\caption{Schematics of the mechanism of the spin elastodynamic motive force. In a ferromagnetic metal, the symmetric part of the deformation couples to magnetization (the magneto-elastic coupling) while the anti-symmetric part couples to electron spin (the spin-vorticity coupling). When a surface acoustic wave is excited in the ferromagnet, these couplings generate two different spatially nonuniform spin dynamics of the magnetization and electron spins. With the s-d coupling between magnetization and spins, the unconventional spin motive force is induced. In particular, the spin-vorticity coupling allows the spin motive force that does not rely on complicated structures and strong spin-orbit interacting materials.  
		}
		\label{fig1}
	    \end{figure}

	We propose the SMF induced by surface acoustic waves (SAWs) via spin-elastodynamics, which we call ``spin elastodynamic motive force (SEMF)", to overcome conventional limitations.
	Three couplings drive the spin-elastodynamics in a ferromagnetic metal: magneto-elastic coupling (MEC), spin-vorticity coupling (SVC), and s-d coupling (Fig.~\ref{fig1}).
	The MEC has been exploited for driving non-equilibrium spin dynamics with a SAW\cite{dreherSurfaceAcousticWave2012,weiler2012}.
	In the presence of a SAW, the SVC, the coupling of electron spins and the vorticity of the lattice of the moving materials, play a significant role for spin manipulations\cite{matsuoEffectsMechanicalRotation2011,matsuoSpindependentInertialForce2011,matsuoSpinCurrentGeneration2011,matsuoRenormalizationSpinrotationCoupling2013,matsuoMechanicalGenerationSpin2013,matsuoTheoryMechanicalSpin2014,matsuoSpinMechatronics2017}. 
	SVC has attracted much attention because it generates spin current without SOI.
	The spin current generated via SVC has been experimentally observed in spin-current generation by vortices associated with liquid metal flow\cite{takahashiSpinHydrodynamicGeneration2016,takahashiGiantSpinHydrodynamic2020,tabaeikazerooniElectronSpinVorticityCoupling2020,tabaeikazerooniElectricalVoltageElectron2021,matsuoTheorySpinHydrodynamic2017} and by vortex motion of lattice associated with SAWs\cite{kobayashiSpinCurrentGeneration2017,kurimuneHighlyNonlinearFrequencydependent2020,kurimuneObservationGyromagneticSpin2020,tatenoElectricalEvaluationAlternating2020,tatenoEinsteinHaasPhase2021}.
	The effective magnetic field owing to the rotational motion has been directly detected using NMR and NQR\cite{chudoObservationBarnettFields2014,chudoRotationalDopplerEffect2015,hariiLineSplittingMechanical2015,onoBarnettEffectParamagnetic2015,arabgolObservationNuclearBarnett2019,chudoObservationAngularMomentum2021,chudoBarnettFieldRotational2021}.

	Due to the rotational motion of the lattice associated with the SAWs,
	the conduction electrons are subjected to SVC when they are applied to ferromagnetic metals.
	Through the s-d coupling, the conduction electron spins adiabatically follow the direction of the magnetization, and the SVC is modulated by the magnetization's precession motion, which causes the SEMF.
	Simultaneously, the MEC operates on the magnetization to excite the magnetization precession owing to lattice strain dynamics.

	The current study finds a way to get over the SMF's conventional limitations.
	Thus, we concentrate on the SEMF induced by a SAW via SVC, a material-independent interaction.
	We calculate the SEMF up to second-order in lattice displacement via treating the influence of magnetization dynamics as a unitary transformation of the spin space. 
	The SEMF has two mechanisms. 
	The first produces an a.c. electromotive force oscillating the SAW's frequency $\omega$, which is induced by the gradient spin accumulation via the SVC (SV-SMF), and the second produces d.c. electromotive force and a.c. one  oscillating $2\omega$ due to the interplay between the SVC acting on conduction electron spins and the MEC acting on magnetization (SVME-SMF).
	The SVME-SMF generates d.c. electromotive force in the film thickness direction when an out-of-plane magnetic field is applied.
	These findings show that the SVME-SMF causes non-reciprocity of the electromotive force.

	{\it Model}.--- We consider the free-electron system and coupled to the magnetization through the s-d coupling.
	The conduction electron's Lagrangian is expressed by
	\begin{gather}
		\mathcal L_{\text{el}}=\int _{\bm x'} c^{\prime \dagger}_{\bm x't}
		\left[
		i\hbar \partial _t + \frac{\hbar^2}{2m} \nabla ^{\prime 2} -V(\bm x') -J_{\text{sd}} \bm m'\cdot \hat{\bm \sigma} 
		\right]c'_{\bm x't},
	\end{gather}
	where $c^{\prime \dagger}_{\bm x't}$ and $c^{\prime}_{\bm x't}$ are electron creation and annihilation operators, respectively, $\hat{\bm \sigma} =(\hat \sigma ^x,\hat \sigma ^y,\hat \sigma ^z)$ are the Pauli matrices, $\bm m'$ is unit vector of the magnetization, $J_{\text{sd}}$ is exchange splitting constant, and $V$ is potential due to the lattice and impurities.
	When the lattice distortion dynamics are induced, the potentials are modulated as $V(\bm x'-\bm u)$, where $\bm u$ is the displacement vector of the lattice.
	Treating the lattice distortion effect, we perform the local coordinate transformation from the laboratory frame $\bm x'$ to the ``rotate frame" $\bm x=\bm x'-\bm u$.
	The conduction electron's Lagrangian is expressed as $\mathcal L=\mathcal L_{\text{el}} + \mathcal L_{\text{sv}}$, where $\mathcal L_{\text{el}}$ is given by
	\begin{gather}
		\mathcal L_{\text{el}} = \int _{\bm x} c^{\dagger}_{\bm xt} \left[
		i\hbar \partial _t + \frac{\hbar^2}{2m}\nabla ^2 - V(\bm x) - J_{\text{sd}} \bm m\cdot \hat{\bm \sigma}
		\right] c_{\bm xt},
	\end{gather}
	where $c_{\bm xt}=\sqrt{1+\nabla \cdot \bm u}c^{\prime}_{\bm x't}$ and $\bm m^{\prime }=\mathcal R( r_{ij})\bm m$ is the unit vector of the magnetization in the rotate frame with $\mathcal R(r_{ij})$ being the $SO(3)$ rotation matrix and $r_{ij} =\frac{1}{2}(\partial_iu^j-\partial_ju^i)$ being the rotation tensor associated with the coordinate transformation.
	The $\mathcal L_{\text{sv}}$ is the SVC:
	\begin{gather}
		\mathcal L_{\text{sv}} = \frac{\hbar}{4} \int _{\bm x} c^{\dagger}_{\bm xt} \hat{\bm \sigma}  \cdot \bm \Omega c_{\bm xt},
	\end{gather}
	where $\Omega_k=2\epsilon_{ijk} \partial_t r_{ij} $ is the vorticity of the lattice motion.
	Note that the modulation of the kinetic energy and the time derivative appear; however, they are not dominant in this study and are neglected.
	We perform rotational transformation in the spin space $c_{\bm xt}=U(\bm x,t)a_{\bm xt}$ with the direction of magnetization $\bm m$ in the $z$-direction, where $a^{\dagger}_{\bm xt}$ and $a_{\bm xt}$ are the electron creation and annihilation operators, respectively, after the rotational transformation.
	Here the unitary operator $U$ is determined to satisfy $c^{\dagger}\bm m\cdot \hat{\bm \sigma} c = a^{\dagger} \sigma ^z a$.
	The Lagrangian including the SVC after the transformation is given by 
	\begin{gather}
		\tilde{\mathcal L}_{\text{el}}= \int _{\bm x} a^{\dagger}_{\bm xt} \left[ i\hbar\partial _t - \frac{(\bm p +e\hat{\bm A} ^{\text s})^2}{2m} - \tilde V +e \hat A^{\text s}_0 -J_{\text{ex}}\sigma ^z \right]a_{\bm xt},
	\end{gather}
	where $\bm p=-i\hbar \nabla$ is the momentum operator and $\tilde V=U^{\dagger}VU$.
	Here, $\hat A^{\text s}_0$ and $\hat{\bm A}^{\text s}$ are the effective spin scalar and vector potentials, given by $\hat A_0^{\text s} = \frac{i\hbar}{e}U^{\dagger}\partial _t U +\frac{\hbar}{4e}U^{\dagger} (\hat{\bm \sigma} \cdot \bm \Omega)U$,
	and 
	$\hat{\bm A}^{\text s} = -\frac{i\hbar}{e} U^{\dagger} \nabla U$.
	Particularly, the second term of $\hat A^{\text s}_0$ is spin scalar potential induced by the SVC.
	The effective electric field is obtained by $\hat{\bm E}^{\text s}=-\partial _t \hat{\bm A}^{\text s} - \nabla \hat A^{\text s}_0$.
	The total SMF is given by $\bm E_{\text{smf}} = P\text{tr}(\hat \sigma ^z \hat{\bm E}^{\text s}),$	where $P$ is the spin polarization of conduction electrons.
	The total SMF contains conventional SMF and the SEMFs.
	\begin{align}
		\bm E_{\text{smf}} = \bm E_{\text{con}} -\frac{P\hbar}{2e} \nabla (\bm m\cdot \bm \Omega),
	\end{align}
	where the first term $E_{\text{con},i} = \frac{P \hbar }{e} \bm m \cdot \left(\partial_t \bm m \times \partial_i \bm m \right)$ is the conventional SMF, and the second term is the SEMFs, which is dominant in this setup.

	The spin dynamics generated by SAWs under a static magnetic field is determined by
	\begin{align}
		\partial_t \bm m' =  \gamma (\bm h_{\text{eff}} + \bm h_{\text{me}}) \times \bm m' - \alpha  (\partial_t \bm m') \times \bm m',
	\end{align}
	where $\gamma(>0)$ is the gyromagnetic constant,  $\alpha $ is the Gilbert damping constant, and $h_{\text{eff}}$ is a static effective magnetic field containing the static magnetic field and an anisotropy field.
	We assume that the Barnett field is negligible compared to the effective magnetic field due to the MEC, which is given by\cite{chikazumi}
	\begin{align}
		h_{\text{me},i} = -\frac{2}{M_{\text S}}\sum_je_{ij}m_j'\Bigl[ b_1\delta _{ij}+b_2(1-\delta _{ij})\Bigr], 
	\end{align}
	where $e_{ij}=\frac{1}{2}( \partial _iu_j + \partial _j u_i)$ is the strain tensor, $M_{\text S}$ is the saturation magnetization, and
	$b_1$ and $b_2$ are the MEC constants.
	The magnetization processes around $h_{\text{eff}}$ with 
	the resonance frequency of the Kittel modes $\omega _{\text K}$ in the spin-wave approximation.
	The magnetization vector in the rotate frame is given by
	$\bm m=\bm m_0+\delta \bm m+\mathcal O(u^2)$, where $\bm m_0=(\sin \theta_0\cos \phi_0, \sin \theta_0 \sin \phi_0, \cos \theta_0)$ is the static component and $\delta \bm m$ is the dynamical component.
	Note that the magnetization modulation associated with the rotational transformation is negligible because of a higher order of Gilbert damping.
	
	The lattice displacement $\bm u$ with the Rayleigh-type SAW (R-SAW) proceeding in the $x$-direction is given by\cite{cleland}
	\begin{align}
		\bm u = u_0 e^{iq_{\mu}x^{\mu}}
		\left(
		\begin{matrix}
			i\mathfrak{s}_q \left( e^{\kappa_{\text t}z} - \frac{2\kappa_{\text t}\kappa_{\text l}}{\kappa_{\text l}^2+q^2} e^{\kappa_{\text t}z} \right) \\
			0 \\
			\frac{\kappa_{\text l}}{|q|} \left(
			e^{\kappa_{\text l}z} - \frac{2q^2}{\kappa_{\text t}^2+q^2} e^{\kappa_{\text t}z}
			\right)
		\end{matrix}
		\right),
	\end{align}
	where $u_0$ is the deformation amplitude of R-SAW, $\mathfrak{s}_q$ is the sign function of $q$, and $\kappa_{\text l}^{-1}$ and $\kappa_{\text t}^{-1}$ are decay lengths of longitudinal and transverse waves, respectively.
	Here, $q_{\mu}=(-\omega,\bm q)$ is the four-vector with $\omega$ being a frequency of R-SAW and $\bm q=(q,0,0)$ being a wavenumber of a longitudinal wave, and $x^{\mu}=(t,\bm x)$ is the four-vector of time and coordinate.
	The vorticity $\bm \Omega$ is given by
	\begin{align}
		\Omega _y =  \frac{2r_0\omega ^2}{c_{\text R}} \mathfrak{s}_q e^{\kappa_{\text t}z} e^{iq_{\mu}x^{\mu}},
	\end{align}
	where $c_{\text R}=\omega/|q|$ is a R-SAW velocity and $r_0=|u_z(z=0)|$.

		\begin{figure}[tbp]
		\centering
		\includegraphics[width=75mm]{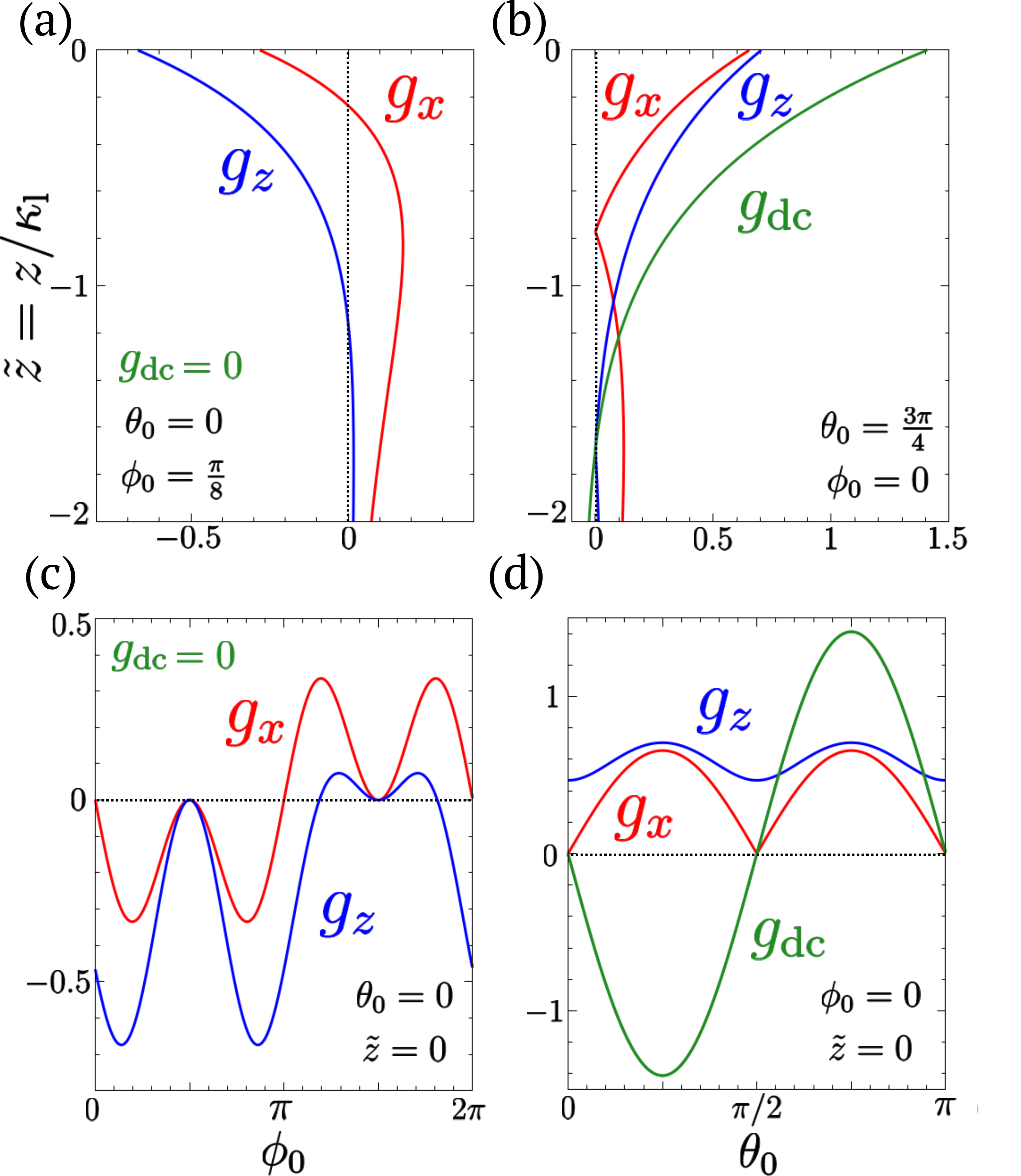}
		\caption{
			The dependence of the SVME-SMF on the depth from the surface (a)(b) and the direction of the applied magnetic field (c)(d).
			The red and blue lines represent the second harmonic component in the $x$-direction and $z$-direction, and
			the green lines represent the d.c. component in the $z$-direction.
			The depth from the surface $\tilde z$ is normalized by the decay constant of the longitudinal wave $\kappa_{\text l}$.
			Here, we assume the isotropic system, and $b_1=b_2$ is satisfied.
			(a) and (c) show the case of an in-plane magnetic field applied ($\theta_0=\frac{\pi}{2}$), and the d.c. component vanishes. 
			(b) and (d) show the case of an out-of-plane magnetic field applied ($\phi_0=0$), and both the second harmonic and d.c. components are induced.
			According to (c), the characteristic non-reciprocity is found in the $z$-direction SMF.
		}
		\label{graph}
	\end{figure}
	
		\begin{table*}[tbp]
		\caption{
			The characteristics of the SEMFs containing the SV-SMF and SVME-SMF.
			The SEMFs do not require any device and material restrictions such as nonuniform device structure, non-collinear magnetization, and strong SOI systems if only the SAW devices are available.
			IP and OP represent the cases where the magnetic field is in the in-plane and out-of-plane directions, respectively.
			The SEMFs 
			produce various types of electromotive forces such as d.c. and a.c. oscillating at $\omega$ and $2\omega$.
		}
		\begin{ruledtabular}
			\begin{tabular}{ccccc}
				Mechanisms & Interaction & Device & IP ($\theta_0=0$) & OP ($\theta_0\neq 0$) 
				\\ \hline \hline
				SVME-SMF & SVC, MEC & \multirow{2}{*}{SAW device} & $2\omega$ & d.c., $2\omega$ 
				\rule[0mm]{0mm}{4mm}
				\\
				SV-SMF & SVC &  & $\omega$ & $\omega$  
				\rule[0mm]{0mm}{4mm}
			\end{tabular}
		\end{ruledtabular}
		\label{list}
	\end{table*}

	{\it Results}.--- The SEMFs contribute to the SMF because the conventional SMF, as is well-known, turns out to vanish in the non-collinear magnetization structure.
	The SMF is given by up to second-order in $u$. 
	\begin{align}
		\bm E_{\text{smf}}=\bm E_{\text{sv}}+\bm E_{\text{s-m}}.
		\label{smf}
	\end{align}
	The first term  $\bm E_{\text{sv}}=-\frac{\hbar}{2e} \nabla (\bm m_0\cdot \bm \Omega)$ is the a.c. component oscillating at $\omega$ due to the gradient spin accumulation induced by the SVC, which we call ``SV-SMF", given by
	\begin{align}
		\left(
		\begin{matrix}
			E_{\text{sv},x} \\ E_{\text{sv},z}
		\end{matrix}
		\right)_{\omega}
		=  \mathcal E_{\text{sv}} \sin \theta_0 \sin \phi_0 e^{\kappa_{\text t}z}
		\left(
		\begin{matrix}
			\sin (q_{\mu}x^{\mu}) \\ -\frac{\kappa_{\text t}}{|q|}\mathfrak{s}_{q} \cos (q_{\mu}x^{\mu})
		\end{matrix}
		\right),
	\end{align}
	where $\mathcal E_{\text{sv}}=\frac{P\hbar r_0}{ec_{\text R}^2} \omega ^3$ a function of $\omega$ with the dimension of an electric field.
	The second term in Eq.~(\ref{smf}) is given by $\bm E_{\text{s-m}}=-\frac{\hbar}{2e} \nabla (\delta \bm m\cdot \bm \Omega)$ from the combination of SVC and MEC, which we call ``SVME-SMF".
	The SVME-SMF contains the second harmonic component oscillating at $2\omega$ and d.c. component. 
	The second harmonic component is given by 
	\begin{align}
		\left(
		\begin{matrix}
			E_{\text{s-m},x}\\
			E_{\text{s-m},z}
		\end{matrix}
		\right)_{2\omega}
		=\mathcal E_{\text{s-m}} \mathfrak{s}_q
		\left(
		\begin{matrix}
			g_x \sin(2q_{\mu}x^{\mu}+\mathfrak{s}_q\Theta_x)\\
			g_z \cos(2q_{\mu}x^{\mu}+\mathfrak{s}_q\Theta_z)
		\end{matrix}
		\right),
	\end{align}
	and the d.c. component is given by
	\begin{align}
		\left(
		\begin{matrix}
			E_{\text{s-m},x}\\
			E_{\text{s-m},z}
		\end{matrix}
		\right)_{\text{dc}}
		= \mathcal E_{\text{s-m}} \mathfrak{s}_q  \left(
		\begin{matrix}
			0 \\
			g_{\text{dc}}
		\end{matrix}
		\right),
	\end{align}
	where $\mathcal E_{\text{s-m}}=\frac{P\hbar\gamma b_1}{2e\omega _{\text K}\alpha M_{\text s}} \frac{u_0 r_0}{c_{\text R}^3} \omega ^4$ is the function of $\omega$, and
	$g_{x/z}$, $g_{\text{dc}}$ are dimensionless functions of $z$, $\theta_0$, $\phi_0$, and $\mathfrak s_q$.
	The dependence of the SVME-SMF on the depth from the surface and direction of the applied magnetic field are shown in Fig.~\ref{graph}.
	According to Fig.~\ref{graph}(c), the second harmonic component on the $z$-direction has characteristic non-reciprocity. 
	The interplay causes this among the three effects: the rotation acting on the electrons via the SVC, the strain acting on the magnetization via MEC, and the s-d coupling\cite{sasaki2017,mingran2018,tateno2020,puebla2020, mingran2020}.
	Note that $\Theta_{x/z}$ are dimensionless functions representing phase shift and turn out to vanish in $\theta_0=\frac{\pi}{2}$.

	{\it Discussion}.---
	Let us estimate the magnitude of the SV-SMF $\mathcal E_{\text{sv}}$ and the SVME-SMF $\mathcal E_{\text{s-m}}$ in polycrystalline nickel with strong ME coupling. 
	When the R-SAW with frequency $f=10$GHz is applied, SV-SMF is estimated as $\mathcal E_{\text{sv}}\sim  5.01\times 10^{-2}$V/m, and SVME-SMF is estimated as $\mathcal E_{\text{s-m}}\sim 5.25\times 10^{-3}$V/m.
	This estimation suggests that an observable voltage, about nV order, can be induced for a distance of about $\mu$m.
	Here, longitudinal wave velocity is $c_{\text l}=6.04\times 10^3$m/s, transverse wave velocity is $c_{\text t}=3.00\times 10^3$m/s \cite{rikanenpyo}, R-SAW velocity is $c_{\text R}=2.80\times 10^3$m/s \cite{landau}, saturation magnetization is $M_{\text s}=0.61$ T \cite{seberinoMicromagneticsLongFerromagnetic1997}, damping constant is $\alpha =4.5\times 10^{-2}$ \cite{walowskiIntrinsicNonlocalGilbert2008}, gyromagnetic ratio is $\gamma =2.41\times10^5$ mA$^{-1}$s$^{-1}$ \cite{meyerExperimentalValuesFe1961}, MEC constant is $b_1=b_2=9.5$MJ/$\text m^3$, and lattice displacement is $u_0= 3.5\times 10^{-12}$m, which is 1\% of the nickel lattice constant.

	Surprisingly, the current processes in typical ferromagnetic materials with simple structures, such as nickel, can induce the SMF.
	Complex device structures or non-collinear magnetization structures are required in conventional SMFs.
	A Rashba device with spatial inversion symmetry breaking, a gate voltage modulation, and a strong SOI material are all material constraints for SOI-induced SMFs.
	Because there are no restrictions on device structures or materials, the SEMFs enable technical applications, such as microfabrication on magnetic materials.
	The SEMFs simultaneously generate the d.c. and a.c. components oscillating at $\omega$ and $2\omega$, and the current mechanisms are summarized in Table~\ref{list}.

	{\it Conclusion.---}
	To address the limits of typical SMF processes, we theoretically analyzed the SEMF, which is the SMF induced by the R-SAW via SVC.
	We discovered that the SEMF has two mechanisms.
	The first is SV-SMF generates the first harmonic electromotive force due to gradient spin accumulation generated by SVC, and the second is SVME-SMF generates d.c. and second harmonic electromotive force due to the combination of SVC and MEC.
	Particularly, the d.c. SMF is induced under an applied out-of-plane magnetic field.
	The results suggest that the second harmonic component on the $z$-direction has characteristic non-reciprocity.
	Our estimation suggests that the electric voltage induced by the SEMFs is detectable in polycrystalline nickel.
	The SEMFs can be generated in ferromagnet without restrictions on device structure or materials if only the SAW device is available.
	Therefore, the SEMFs are expected to expand the range of the SMF applications greatly.

	We would like to thank Y. Nozaki, K. Yamanoi, T. Horaguchi,  J. Fujimoto, and D. Oue for enlightening discussions.
	This work was partially supported by JST CREST Grant No. JPMJCR19J4, Japan.
	TF is supported by JSPS KAKENHI for Grants No. 21K20356.
	MM is supported by JSPS KAKENHI for Grants (Nos. 20H01863 and 21H04565) and the Priority Program of the Chinese Academy of Sciences, Grant No. XDB28000000.

\end{document}